\documentclass{ijuc}
\usepackage[pdftex]{graphics}
\usepackage[dvips]{epsfig}
\usepackage[latin1]{inputenc}
\usepackage[T1]{fontenc}

\usepackage{amssymb, amsmath, amsfonts, latexsym, times, cite, verbatim}
\usepackage{booktabs}
\usepackage{mathrsfs}
\usepackage{amsthm}

\usepackage{algorithm}
\usepackage{algorithmic}

\begin{document}

\title{Green Wireless Sensor Networks with Wireless Power Transfer}

\author{Qiao Li\inst{1}\email{liqiao1989@bupt.edu.cn}
\and Yifei Wei\inst{1}\email{weiyifei@bupt.edu.cn}
\and Mei Song\inst{1}\email{songm@bupt.edu.cn}
\and F. Richard Yu\inst{2}\email{Richard.Yu@carleton.ca}
}

\institute{School of Electronic Engineering, Beijing University
of Posts and Telecommunications, P.R. China
\and
Department of Systems and Computer Engineering,
Carleton University, Canada
}

\maketitle

\begin{abstract}
An energy cooperation policy for energy harvesting wireless sensor networks (WSNs) with wireless power transfer is proposed in this paper to balance the energy at each sensor node and increase the total energy utilization ratio of the whole WSNs. Considering the unbalanced spatio-temporal properties of the energy supply across the deployment terrain of energy harvesting WSNs and the dynamic traffic load at each sensor node, the energy cooperation problem among sensor nodes is decomposed into two steps: the local energy storage at each sensor node based on its traffic load to meet its own needs; within the energy storage procedure sensor nodes with excess energy transmit a part of their energy to nodes with energy shortage through the energy trading. Inventory theory and game theory are respectively applied to solving the local energy storage problem at each sensor node and the energy trading problem among multiple sensor nodes. Numerical results show that compared with the static energy cooperation method without energy trading, the Stackelberg Model based Game we design in this paper can significantly improve the trading volume of energy thereby increasing the utilization ratio of the harvested energy which is unevenly distributed in the WSNs.

\end{abstract}


\section{Introduction}
Recently, energy efficiency has become a hot research topic in wireless networks \cite{XYJL12,BYC12,XYJ12,YZX11,BY14,BYY15,WYS10}. Particularly, energy management is always important for wireless sensor networks (WSNs) due to the limited battery capacity of senor node. A viable approach for sustainably powering the WSNs is to harvest energy from the environment, such as solar, vibrations, thermal, etc. \cite{cite1} \cite{cite2}. For the energy harvesting WSNs, the difference between the available environment power and the power consumption through the network is the key challenge. To use the harvested energy efficiently, the energy  aware task allotment, (i.e., the task distribution among nodes is adapted to the detailed characteristics of environmental energy availability) has been researched extensively in recent years \cite{cite3} \cite{cite4}. However, with the emerging research hot spot of wireless power transfer (WPT) \cite{cite5} \cite{cite6} and simultaneously wireless information and power transfer (SWIPT) \cite{cite7}-\cite{cite11}, the energy management problem of energy harvesting WSNs should be rethought from the perspective of \emph{energy cooperation}.

Energy cooperation method among wireless network nodes powered by renewable energy is studied in \cite{cite12}, where the authors determine energy management policies that maximize the system throughput within a given duration using a Lagrangian formulation and the resulting Karush-Kuhn-Tucker (KKT) optimality conditions. In \cite{cite13}, base stations in coordinated multi-point enabled cellular networks are equipped with energy harvesting devices to provide renewable energy and employ smart meters and aggregator to enable both two-way information and energy flows
with the smart grid. The renewable energy cooperation among
base stations is formulated as a convex optimization problem. Different from the energy cooperation problem in renewable energy powered cellular networks and wireless local area networks in which the aim of energy cooperation is usually optimize the network throughput or increase the performance of the total network and the energy cooperation is usually a optimization problem, the energy cooperation in WSNs is a equilibrium problem because information from each node is equally important for us to surveillance environment. Considering the spatio-temporal properties of the energy supply across the deployment terrain of energy harvesting WSNs and the dynamic traffic load at each sensor node, the aim of energy cooperation among multiple sensor nodes is to balance the energy and traffic at each sensor node and to improve the overall energy utilization ratio of the WSNs.

In this paper we propose the energy cooperation policy for energy harvesting WSNs with WPT, through the local energy storage at each node and energy trade among multiple nodes. A combination of inventory theory and game theory is applied to energy cooperation among sensor nodes for the first time to guarantee the energy supply at each sensor node and simultaneously increase the total energy utilization ratio. The structure of this paper is as follows. Section 2 presents the system model which includes the network model, and the energy supply and demand model at each sensor node. Section 3 formulates the local storage of energy at each sensor node as an inventory problem, and formulates the energy cooperation problem among nodes as a game model. Section 4 solves the inventory problem using inventory theory and designs the Cournot Model based Game and Stackelberg Model based Game to solve the energy trading problem among multiple sensor nodes. Section 5 describes the algorithm flow of the proposed games in detail through a case study and discusses the numerical simulation results. Finally, section 6 concludes the works of this paper.

\begin{figure}
\begin{center}
\scalebox{0.5}{\includegraphics[width=16cm,height=12cm]{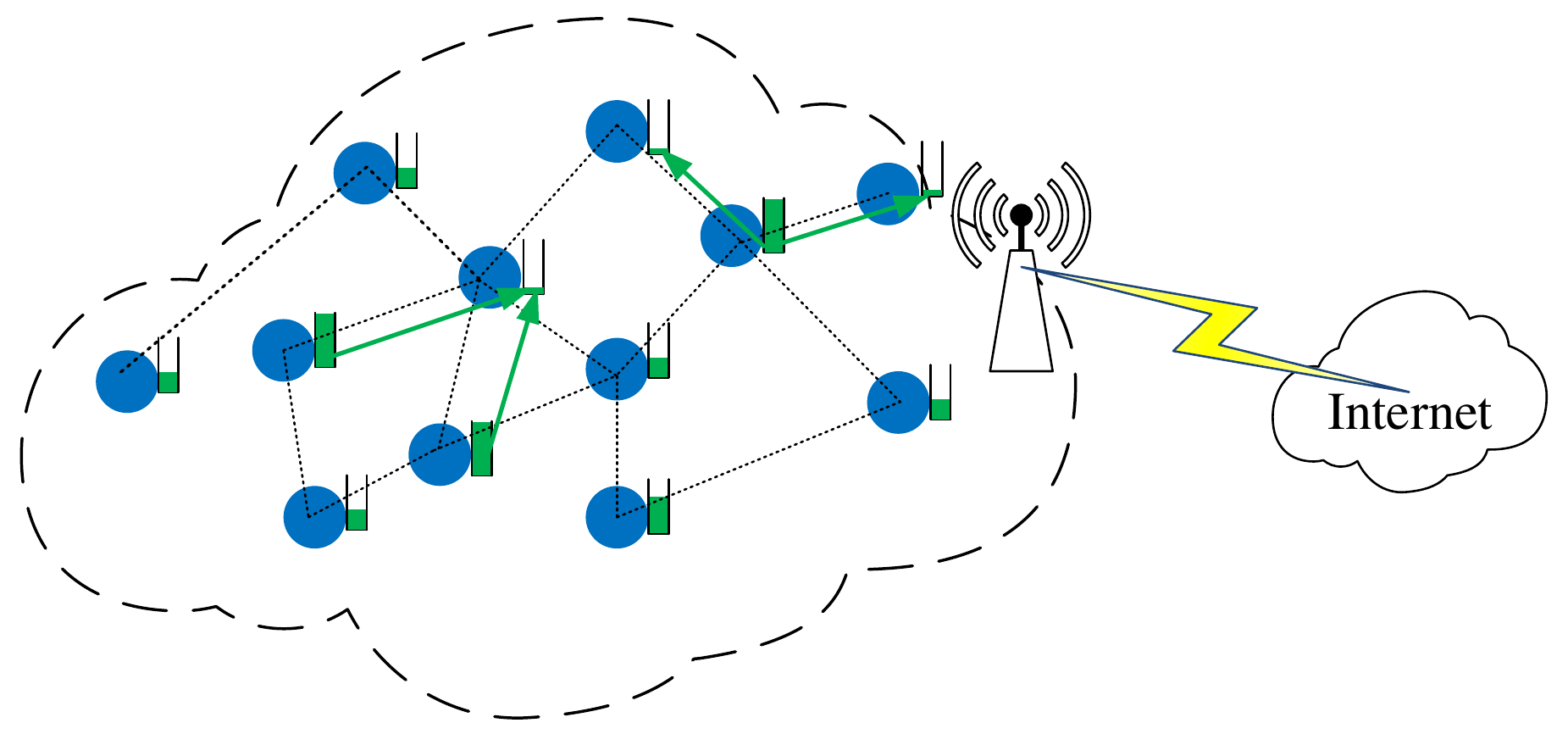}}
\end{center}
\caption{Wireless sensor network with  wireless power transfer.}
\label{fig-eg1}
\end{figure}
\section{System model}
In this section, we present the system model which includes network model, and the energy supply and demand model of each sensor node.

\subsection{Network Model}
Each sensor node equipped with energy harvesting devices collects ambient energy, such as solar power, microbial fuel cells, vibrations and acoustic noise, converses the energy into electrical form, and stores the energy into a rechargeable battery. As shown in Figure 1, energy cooperation among geographically distributed nodes is implemented through the wireless power transfer from one node to the other. Sensor node in the network is either a renewable energy supplier or an energy demander. For simplicity, mobility \cite{YL01} and handover \cite{MYL04,YK07} are not considered in this paper. Each sensor node helps point to point packet delivery through routing packets from its neighbor nodes\cite{cite14}-\cite{cite16}. The radio channel which characterizes the propagation of radio signals among the nodes is modeled as the band-limited Gaussian channel with power constraint \cite{LYH10,YHT10}.

\subsection{Energy Supply: Cooperation based Local Storage}
Energy stored at node $i$ (denoted by $N_i$) and time $t$ is, $S_i(t)\geq 0$. $S_i(t)$ is in general a stochastic process, consisting of its own harvesting energy denoted by $H_i(t)$, and cooperative energy transferred from other node $j$, $N_j$, denoted by $C_{ji}(t)$. The value of $C_{ji}(t)$ can be positive, representing an energy deficit at $N_i$, or be negative, representing a surplus of energy at $N_i$.
\[
S_i(t)= H_i(t)+C_{ji}(t),  \{i, j \in \mathbb{Z}^+; i \neq j\}
\]

For $N_i$, the amount of $H_i(t)$ and $S_i(t)$ is real-timely monitored and recorded.






\subsection{Energy Demand: Data Packets Arrival and Energy Consumption Model}
 The number of data packets arriving at each sensor node has been dynamically changing not only in the time domain but also in the spatial domain. Data packets, which we call traffic load interchangeably in this paper, arriving at one node for a long period of time has great uncertainty, however, during a certain period of time, the arrival process of traffic load can be regarded as Poisson process. Assuming that the arrival rate of data packet at $N_i$ during time period $[t,t+\tau]$ is $\mu$ ($\mu \geq 0$ ), the number of data packet arriving during $[t,t+\tau]$ is a random variable $k$ that subjects to Poisson distribution with parameter $\mu \times \tau$ which means the expected value of traffic load which we call traffic quantity. The probability distribution function (p.d.f.) of $k$ is expressed as:
\[
p_{k}(\tau)=\frac{(\mu \times \tau)^k}{n!}e^{-\mu\tau}, k={0,1,\ldots,n}. \{n \in \mathbb{Z}^+\}
\]
The energy demand of node $N_i$ at time $t$: $D_i(t)$, is also Poisson process, since the energy consumed at each node is primarily used to transmit data packets. For node $N_i$, when the number of data packet arriving during $[t,t+\tau]$ is $k$, the energy demand, $d=d_{k}=a \times k$ is a random variable also obeying Poisson distribution. Here, $a$ represents the average energy consumption of per unit data packet. Denote the probability of $d=d_{k}$ as $p(d)=p_{d_{k}}(\tau)$ then we get,
\begin{equation}
p(d)=p_{d_{k}}(\tau)=p_{k}(\tau)=\frac{(\mu \times \tau)^k}{n!}e^{-\mu\tau}
\end{equation}

\[
s.t. \left\{ \begin{array}{ll}
0 < d_k < d_{k+1} \\
\sum_{k=0}^\infty p_k(\tau)=1  \\
\end{array} \right.
\]

When the arriving rate of data packet $\mu$ changes along with time, the p.d.f. of $p_{d_{k}}(\tau)$ and $p_{k}(\tau)$ changes along with $\mu$ accordingly.

For each node in the WSN, the energy consumed for data packets transmission is determined by two factors which include data rate and link quality. Consider a band-limited Gaussian channel model with power constraint. The achievable data rate at $N_i$ is as follows
\begin{equation}
r_b=b_ilog_2(1+SNR_i)=b_ilog_2(1+\frac{P_i}{b_iN_0})
\end{equation}
where $N_0$ denotes the power spectral density of the additive white Gaussian noise, $P_i \geq 0$ and $b_i \geq 0$ denote the transmit power and bandwidth from one sensor node to node $N_i$ respectively.
To reach a threshold level of data rate at a sensor node with a specific link quality, the transmit power needed is,
\begin{equation}
P_i=b_iN_0(2^{\frac{r_b}{b_i}}-1)
\end{equation}
The terms energy and power are interchangeably used in this paper, since the the transmitting power at each sensor node is the main part of energy consumption.

\section{Problem Formulation}
Through analyzing the distinctive attribute of the energy cooperation among nodes in WSNs, we formulate the local energy storage procedure as an inventory problem and formulate the energy cooperation process among nodes as a game theory problem.

\subsection{Traffic Aware Energy Inventory}

Inventory theory is a branch of operation research \cite{cite17}. A main reason to develop inventory theory is that, for the supply side it is rarely possible to predict the demand exactly. Inventory serves as a buffer against the uncertain and fluctuation of the demand and keeps a supply of items available in case the item are needed by its customers. An inventory problem consists of four basic elements: demand, supply, inventory strategy and inventory cost. The inventory cost $c$ generally consists of four components: setup cost $c_{se}$, holding cost $c_h$, shortage cost $c_s$ and purchasing cost $c_{pur}$, i.e., $c=c_{se}+c_h+c_s+c_{pur}$.

The goal of inventory control is determining the optimal inventory amount $S$ at an appropriate time point $t$ to satisfy the demand $d$ by using a specific inventory strategy at the lowest inventory cost $c$. There will be two different results of inventory amount $S$ for an inventory strategy:
\begin{itemize}

\item $S < d$: The storage amount is less than demand. The shortage cost need to pay is $(d-S)\times C_S$. Here, $C_S$ represents unit shortage cost.

\item $S \geq d$: The storage amount is more than demand.  The holding cost is $(S-d)\times C_H$. Here, $C_H$ represents unit holding cost.

\end{itemize}

At each sensor node, traffic load leads to the demand for energy and the harvested energy acts as the supply. Each node at WSNs respectively determines the optimal amount of the harvested energy reserved for its own traffic load. Inventory strategy for the harvested energy at each node is a bridge that connects traffic load with energy supply. After briefly introducing the inventory theory, we obtain the optimal distributed management of the harvested energy unbalancedly distributed in WSNs through adopting an appropriate inventory strategy to find the optimal inventory amount of the item that is needed, i.e., the harvested energy. As to inventory cost, holding cost and shortage cost is the most important part we consider, since the setup cost and purchasing cost of the ambient energy are negligible. In this paper, holding cost represents the limited capacity of battery that the higher capacity means the higher cost and shortage cost represents expenses due to the deficient storage of energy, such as the loss of packets. When demand $d$ is a stochastic variable and the p.d.f. of $d$ is known, denoted as $P(d)$ , in the case of only considering holding cost and shortage cost, the expected value of inventory cost $EV(c)$ can be expressed as:
\begin{gather}
EV(c)=C_H\times\sum_{d \leq S}(S-d)P(d)+C_S\times\sum_{d> S}(d-S)P(d)
\end{gather}
The optimal inventory amount of renewable is a value of $S$, to make the inventory cost $EV(c)$ minimum.

\subsection{Energy Cooperation: Energy Trading Method between Energy Supplier and Demander}
After calculating the optimal amount of the harvested energy that should be stored in advance through the inventory process at each node, all the sensor nodes are divided into two categories: energy suppliers, i.e., the energy stored in battery is more than demand, and energy demanders, i.e. the energy stored in battery cannot meet the demand. Each energy demander applies for a certain amount of energy from energy suppliers, and the suppliers want to acquire an income from selling their extra energy. The supplier also needs to consider its selling cost resulting from that the energy sold to the demander will not available to the supplier although the supplier need more energy later. Hence, confronting an energy demander with a certain amount of energy demand, each supplier should adopt proper strategies to determine the appropriate amount of energy to sell, and all suppliers get into a relationship of checks and balances to determine the optimal amount to give and the optimal price to take. It forms the game among all energy suppliers.

When it comes to game theory, games are characterized by a number of players or decision makers who interact, possibly threaten each other and form coalitions, take actions under uncertain conditions, and finally receive some benefit or reward or possibly some punishment or monetary loss \cite{cite18}. The strategic form of a game is typically defined by these three objects:
\begin{enumerate}
  \item the set, $M={0,1,\ldots,m}. \{m \in \mathbb{Z}^+\}$, of players,
  \item the sequence, $A_1,\ldots,A_m$, of strategy sets of the players, and
  \item the sequence, $f_1(a_1,\ldots,a_m), \ldots, f_m(a_1,\ldots,a_m)$, of real-valued payoff functions of
the players.
\end{enumerate}
The game is denoted as $G=\{A_1,\ldots,A_m; f_1, \ldots, f_m\}$. Each energy supplier node $j$, $N_j$, acts as a player, and all energy supplier nodes construct a game. Energy supplier $N_j$ choose a strategy ${a_j}^*$ based on its own situation and other suppliers' decision, to maximize its own payoff function $f_j(a_1,\ldots, {a_j}^*, \ldots, a_m)$. Relative to supplier $N_j$, other $m-1$ suppliers respectively choose their best strategies, we call the game reach the Nash Equilibrium, if the payoff function satisfies the following equation:
\begin{gather}
f_j({a_1}^*,\ldots, {a_j}^*, \ldots, {a_m}^*) \geq f_j({a_1}^*,\ldots, a_j, \ldots, {a_m}^*)
\end{gather}
In other words, each player in the game cannot achieve a better payoff only through changing the strategy itself. In this paper, game among all energy suppliers is one kind of duopoly game with complete information (i.e. for each energy supplier, the history decisions of other suppliers are known). Through finding the Nash Equilibrium solution, the ultimate aim of the game theoretic approach of energy cooperation among all sensor nodes is to improve the selling volume of the harvested energy thereby to improve the utilization ratio of the harvested energy distributed in the whole WSN.

\section{Problem Solution}
In this section, we firstly present the inventory control policy
$(s, S)$ where $s$ and $S$ are levels of inventory quantity, and deduce the solution method of the two parameters. After the inventory calculation, we design the game theoretic energy cooperation mode based on the Cournot Model of Duopoly and Stackelberg Model of Duopoly.

\subsection{Energy Inventory}

 Assume the existing storage amount of the harvested energy of node $N_i$ at time $t$ is $I$. In the inventory control policy $(s, S)$, an order is placed to increase the item's inventory amount to the level $S$ as soon as this inventory amount reaches or drops below the level $s$ \cite{cite19}. In this paper we name $s$ inventory bottom line and $S$ optimal Inventory amount. Inventory process proceeds from one time period to the next, and this cycle repeats. When demand is a stochastic variable changing along with time, parameters $s$ and $S$ also change along with time, denoted as $(s(t),S(t))$. Each node at WSN respectively implements $(s, S)$ policy then obtains the optimal amount of energy reserved for its total traffic load.
\subsubsection{Optimal Inventory Amount $S^*$}
As analyzing in section II, energy demand $d$ during time period $[t,t+\tau]$, is a random variable obeying Poisson distribution that the p.d.f is expressed in Equation(1). The quantity of data packet arriving during $[t,t+\tau]$ takes the discrete value, so that energy demand $d$ also takes discrete value. For simplicity, the value of parameters $s$ and $S$ range in $\left\{d_0, d_1, \cdots, d_n\right\}$. When $S$ equals $d_a$, denote $S$ as $S_a$, $(0 \leq a \leq n)$. There are three cases of the relationship between $S$ and $d$:
\begin{itemize}

\item $S < d$: The amount of energy reserved in the battery is deficient. The energy cannot satisfy the demand and sensor node has so limited energy to route packets that results in packets loss and other network problems, which we call energy shortage cost.

\item $S = d$: It is the optimal inventory amount. Energy reserved in the battery exactly satisfies the demand.
\item $S > d$: Energy reserved in the battery exceeds demand and the redundant energy needs additional battery capacity which leads to holding cost.
\end{itemize}

  At each inventory period, on the basis of Equation(4) we deduce the expected value of inventory cost $C(S)$ for the inventory amount $S$:
\begin{gather}
C(S)=c_{se}+C_{PUR}(S-I)+C_H\times\sum_{d \leq S}(S-d)p(d)\notag\\
+C_S\times\sum_{d> S}(d-S)p(d)
\end{gather}
where $p(d)$ is expressed in Equation(1) and $C_{PUR}$ is the unit purchase cost. Take Equation(1) into Equation(6), we get the expected inventory cost for every specific value of $S$.
\begin{gather}
C(S)=c_{se}+C_{PUR}(S-I)\notag\\
+C_H\times\sum_{a \leq k}(S-d_k)\frac{(\mu \times \tau)^k}{n!}e^{-\mu\tau}\notag\\
+C_S\times\sum_{a> k}(d_k-S)\frac{(\mu \times \tau)^k}{n!}e^{-\mu\tau}
\end{gather}
When $S=S_a=d_a$, we obtain the following derivation,
\begin{gather}
\Delta C(S_a)=C(S_{a+1})+C(S_a)\notag\\
=C_{PUR} \times \Delta S_a + C_H \times \Delta S_a \sum_{d \leq S_a}p(d)-SC \times \Delta S_a \sum_{d > S_a}p(d)\notag\\
=C_{PUR} \times \Delta S_a + C_H \times \Delta S_a \sum_{d \leq S_a}p(d)\notag\\
-C_S \times \Delta S_a (1-\sum_{d > S_a}p(d))\notag\\
=(\sum_{d > S_a}p(d)-\frac{C_S-C_{PUR}}{C_S+C_{PUR}})\times (C_H-C_S)\times \Delta S_a
\end{gather}
Denote $\sum_{d > S_a}p(d)$ as $F(S_a)$, and $\frac{C_S-C_{PUR}}{C_S+C_{PUR}}$ as $F$. Generally, $F$ has a value between 0 and 1 $(0\leq F\leq 1)$. Now Equation(8) is rewritten as:
\begin{equation}
\Delta C(S_a)=(F(S_a)-F)\times (C_S+C_S)\times \Delta S_a
\end{equation}
where $(C_H+C_S)\times \Delta S_a \geq 0$, and $F(S_a)$ monotonically increases along with $a$. The plus or minus characteristic of $\Delta C(S_a)$ is the same with that of $F(S_a)-F$ so that $\Delta C(S_a)$ is also monotonically increasing.

The p.d.f. of $k$ analyzed in section II
shows that the probability of one data packet or $n$ data packets arriving during time period $[t,t+\tau]$, i.e., $p_1(\tau)$ and $p_n(\tau)$, are extremely low. Set $F(S_1)=\sum_{d \leq S_1}p(d)=p_1(\tau)$, due to the very small value of $p_1(\tau)$, then take $F(S_1)=p_1(\tau) < F$. Similarly, take $F(S_{n-1})=1-p_n(\tau) > F$, i.e., $p_n(\tau)<1-F$. Now we get these following equations:
\[
\Delta C(S_1)=(F(S_1)-F)\times (C_H+C_S) \times \Delta S_1
\]
\[
=(p_1-F)(C_H+C_S)\times \Delta S_1<0
\]
\[
\Delta C(S_{n-1})=(F(S_{n-1})-F)\times (C_H+C_S) \times \Delta S_{n-1}
\]
\[
=(1-p_n-F)(C_H+C_S)\times \Delta S_{n-1}>0
\]
We have known that $\Delta C(S_a)$ is monotonically increasing along with $a$, and now we conclude that the value of $\Delta C(S_a)$ increases from a minus value to a plus one. Accordingly, the value of $C(S_a)$ firstly increase and then decrease. Therefore, a value $S^*=S_{a^*}=d_{a^*}$ is subsistent to make $C(S_a)$ minimum. Here, $a^*$ is a value in in $\left\{0, 1, \cdots, n\right\}$. The value $S^*$ definitely makes these following relational expressions simultaneously valid:
\begin{itemize}

\item $\Delta C(S_{a^*-1})<0$, according to Equation(10): $F(S_{a^*-1}-F)<0$, i.e.,$F(S_{a^*-1})<F$.

\item $\Delta C(S_{a^*})\geq 0$, according to Equation(10): $F(S_{a^*}-F) \geq 0$, i.e.,$F(S_{a^*}) \geq F$.
\end{itemize}

The optimal inventory amount $S^*$ that makes the expected value of inventory cost minimum can be obtained from the following equation:
\begin{equation}
\sum_{d \leq S_{a^*-1}}p(d)<F=\frac{C_S-C_{PUR}}{C_S+C_H} \leq \sum_{d \leq S_{a^*}}p(d)
\end{equation}

Two extreme situations:
\begin{itemize}

\item $F(S_1)=p_1 \geq F$, i.e., $\Delta C(S_1)>0$. In this case, for an arbitrary value of $a$$(a=0,1,2, \cdots, n-1)$ the relational expression: $\Delta C(S_a)>0$ is always established, that is $S=S_1=d_1$. Additionally, the relational expression:$F<F(S_1)=\sum_{d \leq S_{a^*}}p(d)$ demonstrates that the value of $p(d_1)$ is very large. In other words, the probability of a very less demand for energy is very high, and this corresponds to the situation of very low traffic load at a sensor node.

\item $F(S_{n-1})=1-p_n<F$, i.e., $\Delta C(S_{n-1})>0$. In this case, for an arbitrary value of $a$$(a=0,1,2, \cdots, n-1)$ the relational expression: $\Delta C(S_a)<0$ is always established, that is $S=S_n=d_n$. Additionally, the relational expressions $F(S_{n-1})<F$ and $F<1=F(S_n)$ demonstrate that the value of $p(d_n)$ is very large. That is to say, the probability of a great demand for energy is very high, and this corresponds to the situation of very high traffic load at a sensor node.
\end{itemize}
\subsubsection{Inventory Bottom Line $s$}
The optimal inventory amount $S^*$ has been derived in previous subsection. When the existing storage amount $I$ reaches the level $s$, the expected value of cost resulting from not increasing inventory should be less than that of increasing inventory to $S^*$, expressed by the relational expression:
\[
\sum_{d \leq s}C_H(s-d)p(d)+\sum_{d > s}C_S(d-s)p(d)
\]
\[
\leq c_{se}+C_S(S^*-s)+\sum_{d \leq S^*}C_H(S^*-d)p(d)+\sum_{d > S^*}C_S(d-S^*)p(d)
\]
i.e.,
\begin{gather}
C_{PUR}\times s+\sum_{d \leq s}C_H(s-d)p(d)+\sum_{d > s}C_S(d-s)p(d)\notag\\
\leq c_{se}+C_{PUR}\times S^*+\sum_{d \leq S^*}C_H(S^*-d)p(d)+\sum_{d > S^*}C_S(d-S^*)p(d)
\end{gather}
Equation(11) is obviously valid, when $s=S^*$, so that there must be at least one value of $s$ to make Equation(11) established. Since the inventory item for each sensor node at WSN is the harvested energy which is harvested freely at the sensor node itself, the setup cost $c_{se}$ and unit purchasing cost $C_{PUR}$ are negligible. Let $C_{PUR}$ and $c_s$ in Equation(11) equal zero, we find that in the scenario described in this paper, the order point $s$ is exactly the optimal inventory amount $S^*$. That is to say, for each sensor node, an order is placed to increase the inventory amount of energy to the level $S^*$ if the inventory amount drops below the level $S^*$ at each set time period.

\subsection{Energy Trading Method: A Game Theoretical Approach}
We first discuss the existence of the Nash Equilibrium solution in more general games, more details can be found in literature \cite{cite18} and its references.
\newtheorem{thm}{Theorem}
\begin{thm}
(Nash 1950): In the n-player normal-form game $G=\{A_1,\ldots,\notag\\A_m; f_1, \ldots, f_m\}$, if $n$ is finite and $A_i$ is finite for every $i$ then there exists at least one Nash equilibrium, possibly involving mixed strategies.
\end{thm}
A more universal method to determine the existence of Nash equilibrium solution \cite{cite18} is to verify whether the game process meets the following conditions:
\begin{enumerate}
  \item the number of player $n$ is finite;
  \item the strategy set (i.e., action set) $\{A_i, \ldots\}$ is bounded closed and convex set.
  \item the payoff function in the action set is continuous and quasi concave.
\end{enumerate}
The first two conditions are easy to be satisfied. To achieve the Nash equilibrium solution, the key point is designing appropriate payoff function for players. In what follows, strictly according to the requirements stated above and the network feature of WSN, we design the payoff function for both energy supplier and demander node at the energy harvesting WSN. The games proposed in this paper is based on the classical Cournot Model of Duopoly which is Static game with complete information, and the classical Stackelberg Model of Duopoly which is Dynamic Games with Complete Information.

When a sensor node (which is an energy demander) requests for a certain amount of energy, the energy supplier nodes execute the game theory process and provide an appropriate amount of energy. If the amount of energy each supplier determines to sell stays the same for three times in succession of the games, it indicates achieving the Nash Equilibrium solution. An energy cooperation process consists of the following steps:
\begin{enumerate}
  \item Sensor node which is energy demander makes a request for buying energy and broadcast this request to energy suppliers.
  \item Energy supplier nodes receive the request and each energy supplier get into the decision making state to determine the amount of energy to sell through a game theoretic approach.
  \item Stop the game process when the amount and price of energy provided by energy suppliers reach the steady value, i.e., achieve the Nash Equilibrium solution.
  \item Each energy supplier transmit the energy with the amount determined by step 2 to the energy demander node.
  \item The energy demander node receive the energy transmitted from energy suppliers.
\end{enumerate}

\subsubsection{Cournot Model based Game}
Cournot Model of Duopoly is a kind of static game with complete information. The form of Cournot Model of Duopoly is as follows: first the players simultaneously choose actions; then the players receive payoffs that depend on the combination of actions just chosen. At each move in the game the player with the move knows the full history of the play of the game thus far, and each player's payoff function is common knowledge among all the players. Considering the particular attributes of the energy harvesting WSN, we design the game among energy suppliers based on the Cournot Model. The payoff function of energy supplier node is,
\begin{equation}
f_s(p_i)=qp_i-C(p_i)
\end{equation}
where $p_i$ is the amount of energy sold from energy supplier node $N_i$, $q$ is the price of selling the energy, and $C(p_i)$ is the cost function of selling $p_i$ amount energy.

The price $q$ of energy is determined by the total amount of energy sold by all the suppliers. The precise relationship between the total sell-through and the price of energy can be deduced from the payoff function of the energy demander. Referring to literature \cite{cite20}, we design the payoff function,
\begin{equation}
f_d(p)=\sum_{i=1}^{M} p_ik_i^{(d)}-\frac{1}{2}(\sum_{i=1}^{M} p_i^2+2\sum_{i\neq j}^{M} p_ip_j)-\sum_{i=1}^{M} qp_i
\end{equation}
where $k_i^{(d)}$ is the energy efficiency of demander, defining as the ratio of the data rate of the energy demander node to the power used for transmitting these packets. The energy efficiency can be deduced from Equation (3):
\begin{equation}
k_i^{(d)}=\frac{r_b}{P_i}=\frac{r_b}{b_iN_0(2^{\frac{r_b}{b_i}}-1)}
\end{equation}
Equation (13) is a quadratic and quasi concave function, and there must be a maximum value. Taking the derivative of the amount of power $p$ in Equation (13), we achieve the maximum value of payoff of the demander and the relationship between energy price and amount.
\begin{equation}
\frac{\partial {f_d(p)}}{\partial {p_i}}=k_i^{(d)}-\sum_{i=1}^{M} p_i-q=0
\end{equation}
We get the price of energy,
\begin{equation}
q=k_i^{(d)}-\sum_{i=1}^{M} p_i
\end{equation}
Take Equation (14) into (16), we achieve the relationship between energy price and the sales,
\begin{equation}
q=\frac{r_b}{b_iN_0(2^{\frac{r_b}{b_i}}-1)}-\sum_{i=1}^{M} p_i
\end{equation}

Then we propose the selling cost function $C(p_i)$ of the energy supplier. If a supplier decides to sell a certain amount of energy to a demander, the sold energy can not been taken back, although the supplier faces serious shortages of energy later. In other words, suppliers take risks to sell their energy. Based on the impact of the quality of service of the supplier node resulting from selling energy, we design the selling cost function $C(p_i)$.
\begin{equation}
C(p_i)=wD_i(P_i^{req}-k_i^{(s)}\frac{S_i-p_i}{D_i})^2
\end{equation}
where $D_i$ is the traffic quantity of data packets, $P_i^{req}$ is the amount of energy needed, $S_i$ is the amount of harvested energy stored at $N_i$, $k_i^{(s)}$ is the energy utilization ratio of energy supplier node, $k_i^{(s)}=P_i^{req}/ (S_i / D_i)$, $w$ is the weight of cost function. The higher value of $w$ means a longer distance between supplier and demander, and that means a greater cost causing by power transmit loss.

Take Equation (17) and (18) into (12), we achieve the payoff function of the energy supplier node, as follows.
\begin{equation}
f_s(p_i)=(\frac{r_b}{b_iN_0(2^{\frac{r_b}{b_i}}-1)}-\sum_{i=1}^{M} p_i)p_i-wD_i(\frac{P_i^{req}p_i}{S_i})^2
\end{equation}

The game process: Each energy supplier sells energy with the same quality and price. The price fluctuates with the demand. The action of each supplier is selfish an uncooperative, and the game among suppliers is the determination of selling volume of energy. According to other suppliers history data of selling volume, each supplier determines the most suitable selling volume. The suppliers simultaneously choose actions to determine their own selling volume. Then the suppliers receive payoffs that depend on the combination of actions just chosen. After several times games, all energy suppliers get into the balanced state and each supplier achieves the stable equilibrium value of the selling volume of energy. In section 5, through a case study the game algorithm will be introduced in detail.

\subsubsection{Stackelberg Model based Game}
Stackelberg Model of Duopoly is a kind of dynamic games with complete information. The payoff function of energy supplier and demander, and wireless channel model of the Stackelberg Model are the same with that of Cournot Model based Game we analyzed above. The difference is the game process: In the Stackelberg Model based Game, a certain number of suppliers move first and another part of suppliers move second. The detailed algorithm will be presented through a case study in section 5.
\begin{algorithm}[ht]
\caption{Cournot Model based Game}
\begin{algorithmic}[1]

\STATE Initialization: $r_b=40k$ bits, $b_i=10M$ Hz, $N_0=-50$ dBm, $w=0.5$, $D_i=15$, $P_i^{req}/S_i=120uW/160uW$;\
\FOR{$i=1:M$}
\STATE Simplify the payoff function of energy supplier node $N_i$: $f_s(p_i)=(357-\sum_{j=1,j\neq i}^{M} p_j-p_i)p_i-4p_i^2$;\
\STATE Deduce the expression of $p_i$ by solving the partial differential equation: $\frac{\partial {f_s(p)}}{\partial {p_i}}=0$, thus far, $p_i= \frac{1}{10}(357-\sum_{j=1,j\neq i}^{M} p_j)$;\
\STATE Achieve the value of $p_i$ by taking the history data of $p_j$ into the above function.
\ENDFOR

\label{code:recentEnd1}
\end{algorithmic}
\end{algorithm}

\section{case study and simulation}
In this section, we first set the parameters values for our system model and state the game model we proposed in section 4.2 in detail. Then we provide numerical results for evaluating the performance of the energy cooperation policy we proposed based on the game theoretic framework.
\subsection{Case Study}
At sensor node $N_i$, data packet arrives during time period $[t,t+\tau]$ at rate $\mu$. The expected value of traffic load $\mu \times \tau$ that we call traffic quantity is respectively set to 5, 10, 20. One data packet corresponds to one unit energy 1. Neglect setup cost and purchasing cost, for that the harvested energy is free. Two cases are taken into consideration. In the case which the holding cost is lower than shortage cost, we set the following values: unit holding cost $C_H=1$; unit shortage cost $C_S=4$; the existing storage amount $I=2$ units of energy. In the case which the holding cost is higher than shortage cost, we set the following values: unit holding cost $C_H=4$; unit shortage cost $C_S=3$; the existing storage amount $I=2$ units of energy.
\begin{algorithm}[ht]
\caption{Stackelberg Model based Game}
\begin{algorithmic}[1]

\STATE Initialization: $r_b=40k$ bits, $b_i=10M$ Hz, $N_0=-50$ dBm, $w=0.5$, $D_i=15$, $P_i^{req}/S_i=120uW/160uW$;\

\STATE The value of selling volume of each first moving supplier, denoted as $B_x$, is known. The total value of the selling volume of the first $m$ moving suppliers is $\sum_{x=1}^{m}B_x$ and the number of second moving suppliers is set as $n$;\
\STATE Simplify payoff function of the second moving node $N_i$: $f_s(p_i)=(357-\sum_{x=1}^{m}B_x-\sum_{j=1,j\neq i}^{n} p_j-p_i)p_i-4p_i^2$;\
\STATE Deduce the expression of $p_i$ by solving the partial differential equation: $\frac{\partial {f_s(p)}}{\partial {p_i}}=0$, thus far, $p_i= \frac{1}{10}(357-\sum_{x=1}^{m}B_x-\sum_{j=1,j\neq i}^{n} p_j)$;\
\STATE Solve the value of $B_x$ of the first moving suppliers and accumulate the value of the $n$ second moving suppliers: $\sum_{i=1}^{n}p_i=\frac{n(357-\sum_{x=1}^{m}B_x)}{n+9}$;\
\STATE The payoff function of the first moving node: $f_s(B)=(357-\sum_{i=1}^{n}p_i-\sum_{y=1,y\neq x}^{m} B_y-B_x)B_x-4B_x^2$
$=(357-\frac{n(357-\sum_{x=1}^{m}B_x)}{n+9}-\sum_{y=1,y\neq x}^{m} B_y-B_x)B_x-4B_x^2$;\
\STATE Achieve the value of $B_x$ to maximize the payoff by solving the partial differential equation: $\frac{\partial {f_s(B)}}{\partial {B_x}}=0$;\
\STATE Take $B_x$ back into step 4, the value of the selling volume of the first $m$ moving suppliers $p_i$ is solved.

\label{code:recentEnd2}
\end{algorithmic}
\end{algorithm}
The Crossbow Berkeley motes are one of the most versatile
wireless sensor network devices on the market for prototyping
purposes. In this paper, we take the correlated parameters of Berkeley motes as standard to set the parameters values of our model. The operating frequency of Berkeley motes are in ISM (Industrial Scientific Medical) band, either 916.5 MHz or 433 MHz, with a data rate of
40 kilobits per seconds, and having a range of 30 feet to 100 feet.
Set the threshold value of data rate of each sensor node as 40k bits, $r_b=40k$ bits, the bandwidth as 10 MHz, $b_i=10M$ Hz, and the the power spectral density of the additive white Gaussian noise as -50 dBm, $N_0=-50$ dBm. Set the weight of the selling cost function $w$ as $0.5$, i.e., the transmit loss of the wireless power transfer is $50$ percent. $w=0.5$, $D_i=15$, $P_i^{req}/S_i=120uW/160uW$. Take these parameter values into the Cournot Model based Game and the Stackelberg Model based Game, and the algorithm flow of game processes are represented in \textbf{Algorithm 1} and  \textbf{Algorithm 2} respectively. The completely static energy cooperation method in which all the suppliers provide the same amount of energy is presented in \textbf{Algorithm 3} as a comparison.

\begin{algorithm}[ht]
\caption{Static Energy Cooperation Method}
\begin{algorithmic}[1]

\STATE Initialization: $r_b=40k$ bits, $b_i=10M$ Hz, $N_0=-50$ dBm, $w=0.5$, $D_i=15$, $P_i^{req}/S_i=120uW/160uW$;\
\FOR{$i=1:M$}
\STATE Simplify the payoff function of energy supplier node $N_i$: $f_s(p_i)=(357-\sum_{i=1}^{M}p_i)p_i-4p_i^2$. Since each energy supplier node has the same trade strategy, i.e., the selling volume is the same, the payoff function is $f_s(p_i)=(357-Mp_i)p_i-4p_i^2$;\
\STATE Deduce the expression of $p_i$ by solving the partial differential equation: $\frac{d(f_s(p))}{d (p_i)}=0$;\
\STATE Achieve the selling volume of each energy supplier: $p=\frac{357M}{2M+8}$;\
\ENDFOR

\label{code:recentEnd3}
\end{algorithmic}
\end{algorithm}

\subsection{Numerical Simulation and Analysis}

\subsubsection{Optimal inventory amount of energy at each sensor node}
\begin{figure}
\begin{center}
\scalebox{0.5}{\includegraphics[width=16cm,height=12cm]{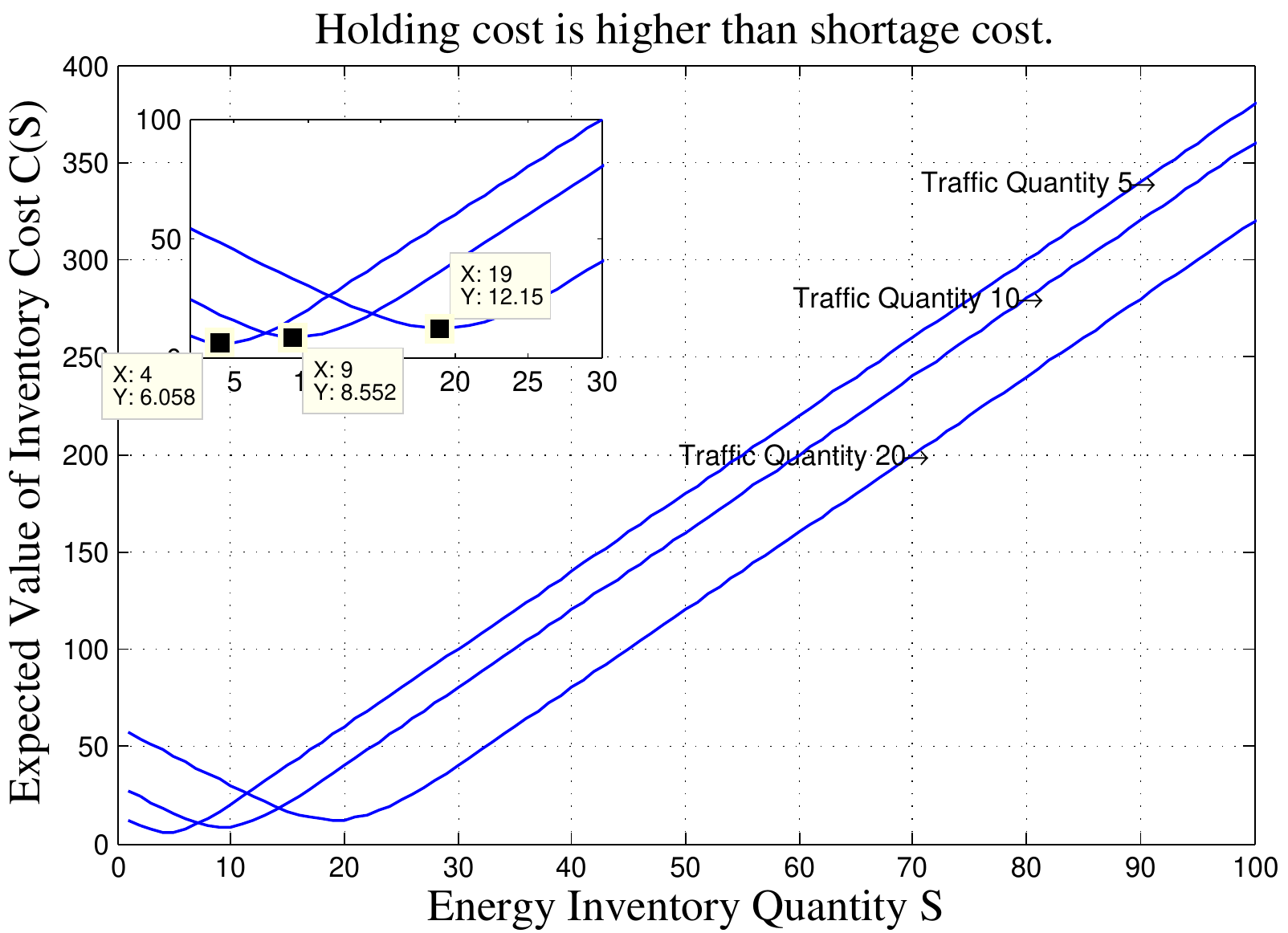}}
\end{center}
\caption{The optimal inventory quantity of energy at the sensor node with a higher holding cost.}
\label{fig-eg2}
\end{figure}

\begin{figure}
\begin{center}
\scalebox{0.5}{\includegraphics[width=16cm,height=12cm]{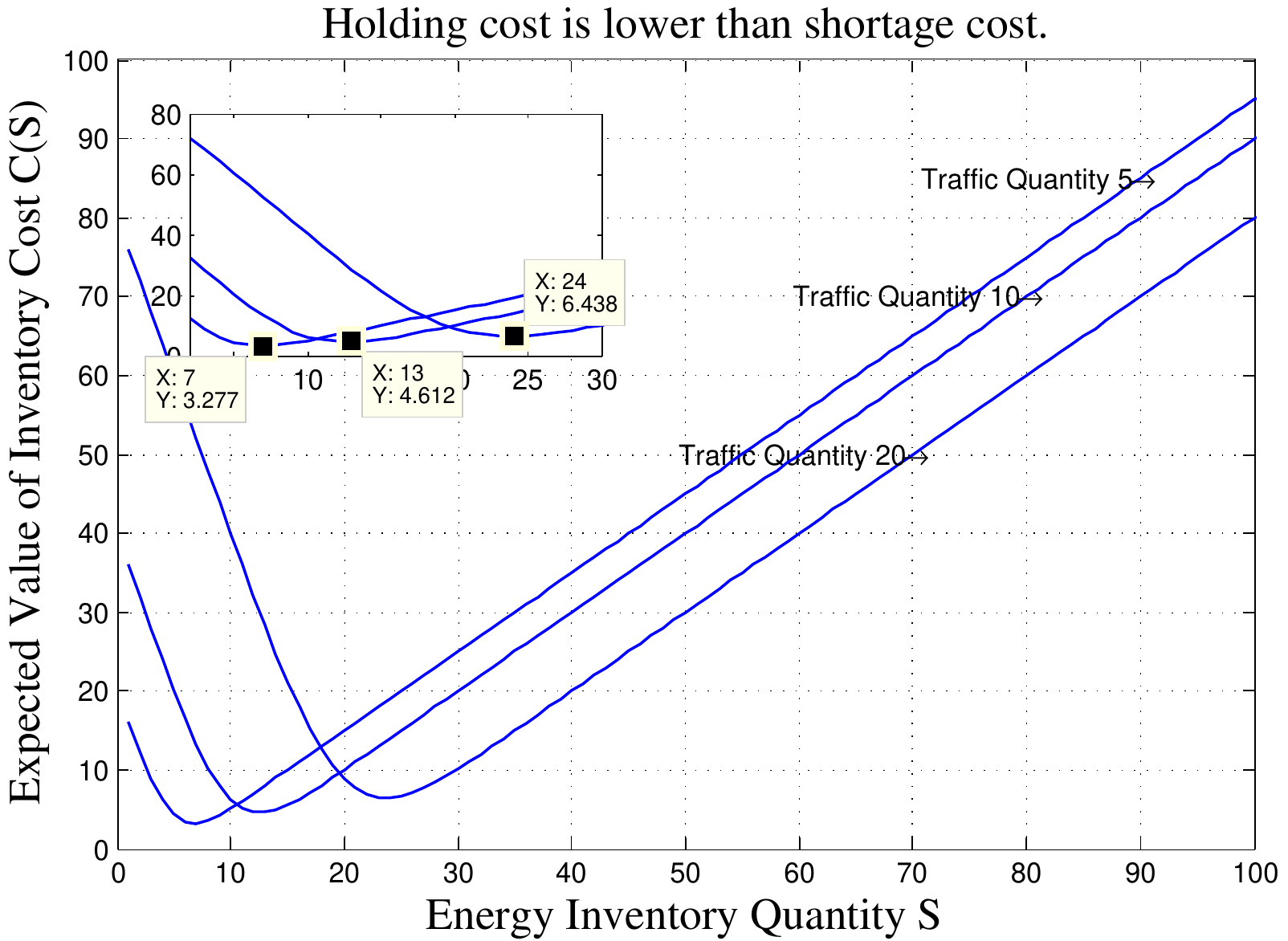}}
\end{center}
\caption{The optimal inventory quantity of energy at the sensor node with a lower holding cost.}
\label{fig-eg3}
\end{figure}
Figure 2 shows the expected value of inventory cost with a higher holding cost of the harvested energy. The optimal inventory quantity of energy makes the expected value of inventory cost at a minimum level. For sensor nodes with different traffic quantity during a time period, their optimal inventory amount of energy and the minimum inventory cost are correspondingly different. As shown in the figure, when the traffic quantity is $5$, $10$ and $20$ units, the optimal inventory quantity of energy is $4$, $9$ and $19$ units respectively, with the minimum inventory cost $6.058$, $8.552$ and $12.15$.

When the holding cost is lower, the optimal inventory amount of harvested energy is shown in Figure 3. It is observed that the optimal inventory quantity of energy is $7$, $13$ and $24$ units respectively, with the minimum inventory cost $3.277$, $4.621$ and $6.438$, when the traffic quantity is $5$, $10$ and $20$ units. Compared with Figure 2, we can see that, with an equal quantity of traffic, lower holding cost means that the optimal inventory amount of harvested energy is higher, and meanwhile the inventory cost is lower. It illustrates that, for each sensor node, the cooperation based inventory amount of harvested energy
depends on the storage capacity and cost, and the traffic quantity.

\subsubsection{Game theoretical approach to energy cooperation among sensor nodes}
\begin{figure}
\begin{center}
\scalebox{0.5}{\includegraphics[width=16cm,height=12cm]{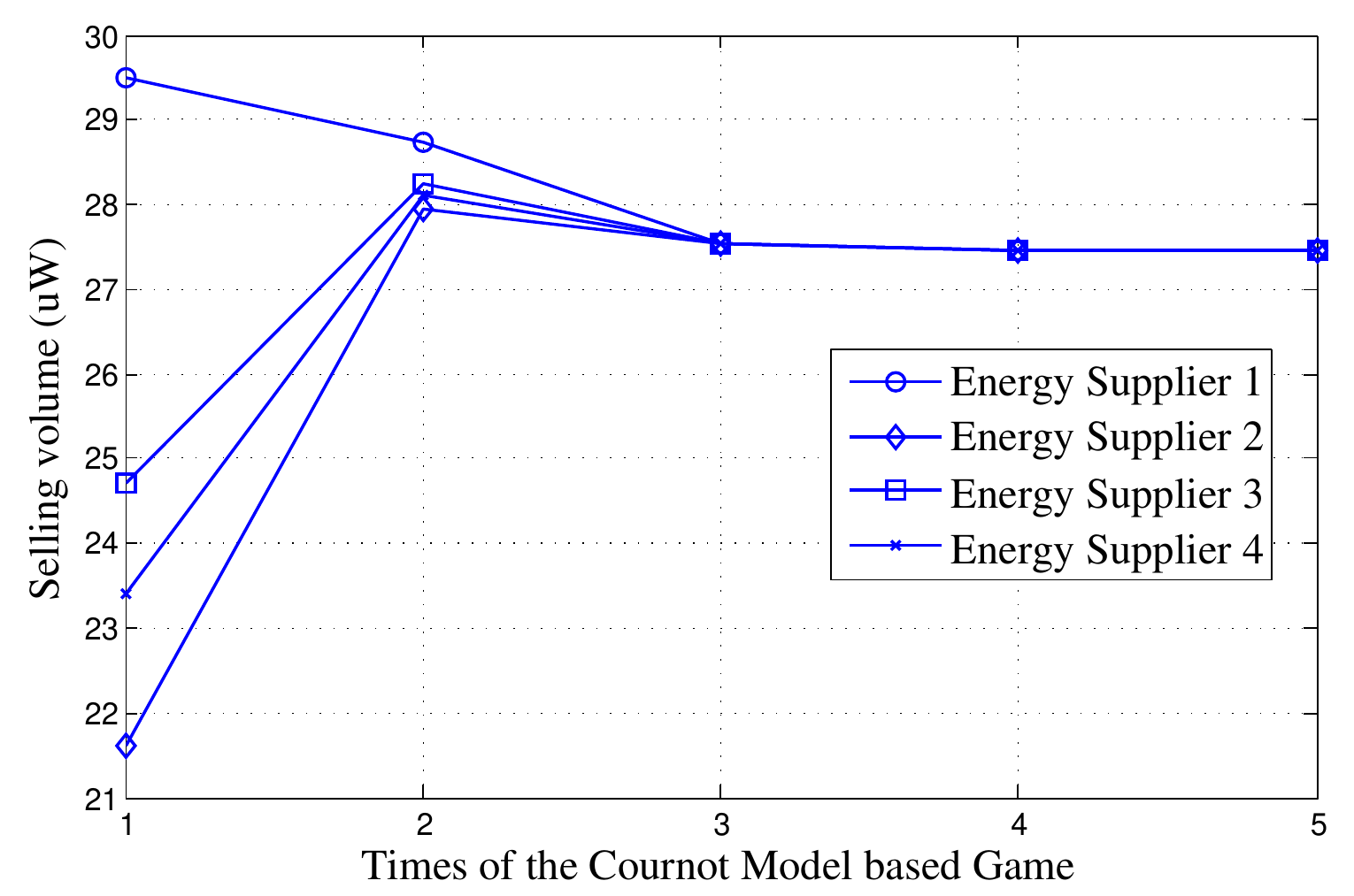}}
\end{center}
\caption{Game process of achieving the Nash equilibrium solution of the Cournot Model based Game.}
\label{fig-eg4}
\end{figure}

\begin{figure}
\begin{center}
\scalebox{0.5}{\includegraphics[width=16cm,height=12cm]{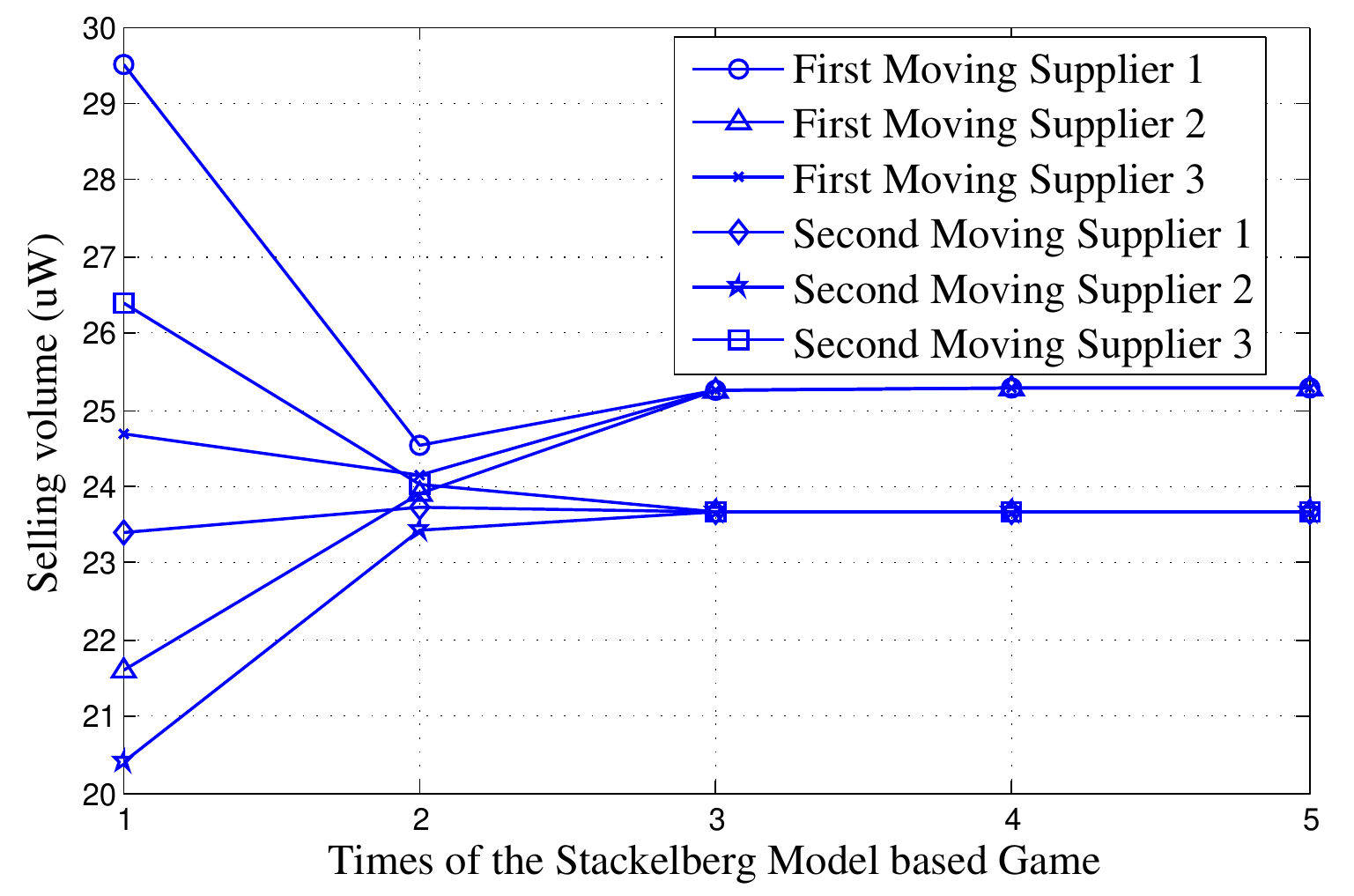}}
\end{center}
\caption{Game process of achieving the Nash equilibrium solution of the Stackelberg Model based Game.}
\label{fig-eg5}
\end{figure}
The process of achieving the Nash equilibrium solution of the Cournot Model based Game is shown in Figure 4. Four energy suppliers are in this game, and the history selling volume of energy of each supplier is set as $29.5 uW$, $21.6 uW$, $24.7 uW$ and $23.4uW$. As shown in the figure, after $5$ times game, the equilibrium is achieved at the sixth time. The process of obtaining the Nash Equilibrium solution of the Stackelberg Model based Game is shown in Figure 5. Six energy suppliers are in the game with three first moving suppliers and three second moving suppliers. The history decision is set as $29.5 uW$, $21.6 uW$, $24.7 uW$, $23.4uW$, $20.4uW$ and $26.4uW$ respectively. After 5 times game, the three first moving suppliers come to an equilibrium value and the second moving suppliers come to the other equilibrium value.
\begin{figure}
\begin{center}
\scalebox{0.5}{\includegraphics[width=16cm,height=12cm]{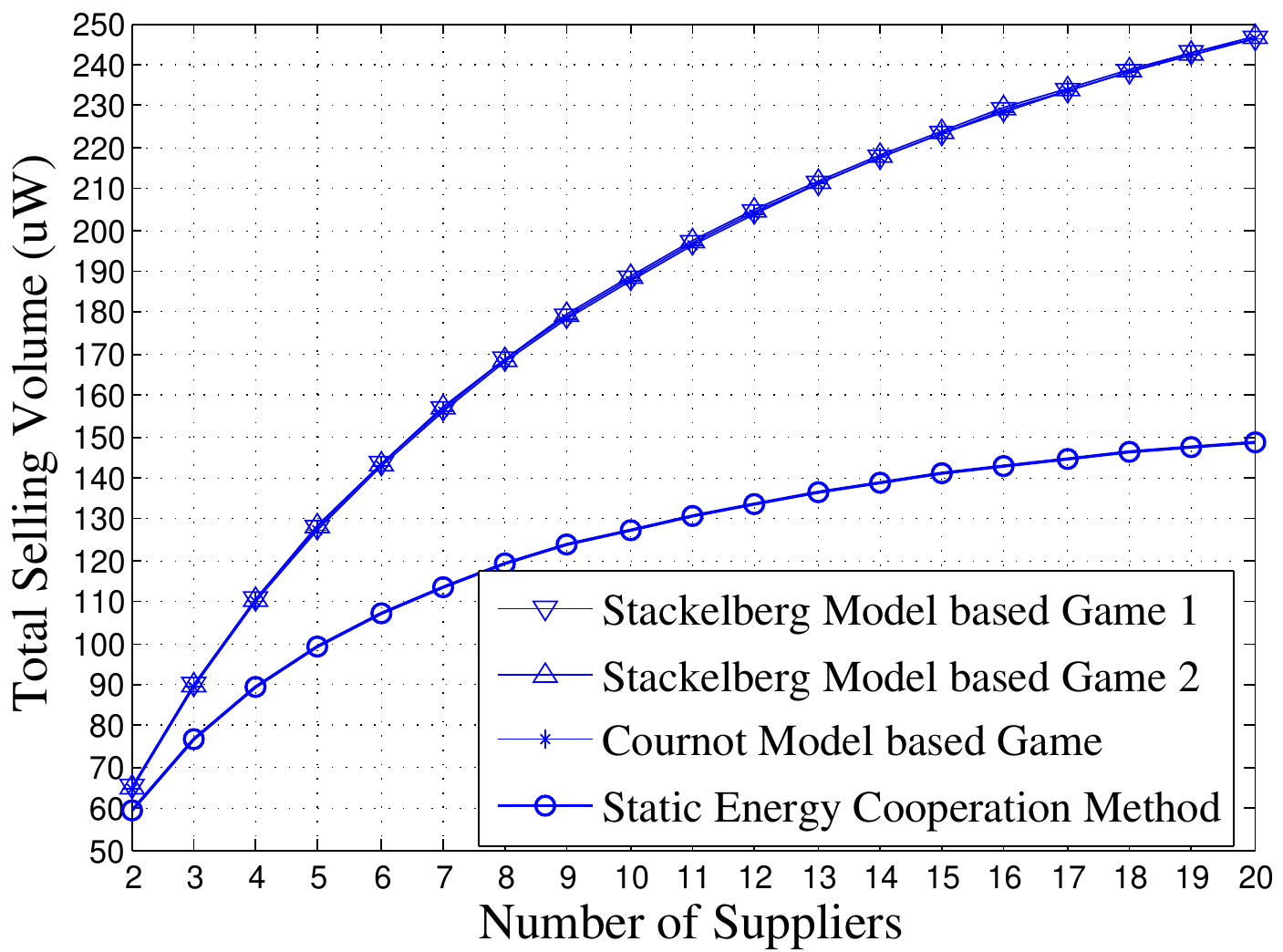}}
\end{center}
\caption{Total selling volume of energy with different energy trading method.}
\label{fig-eg6}
\end{figure}

\begin{figure}
\begin{center}
\scalebox{0.5}{\includegraphics[width=16cm,height=12cm]{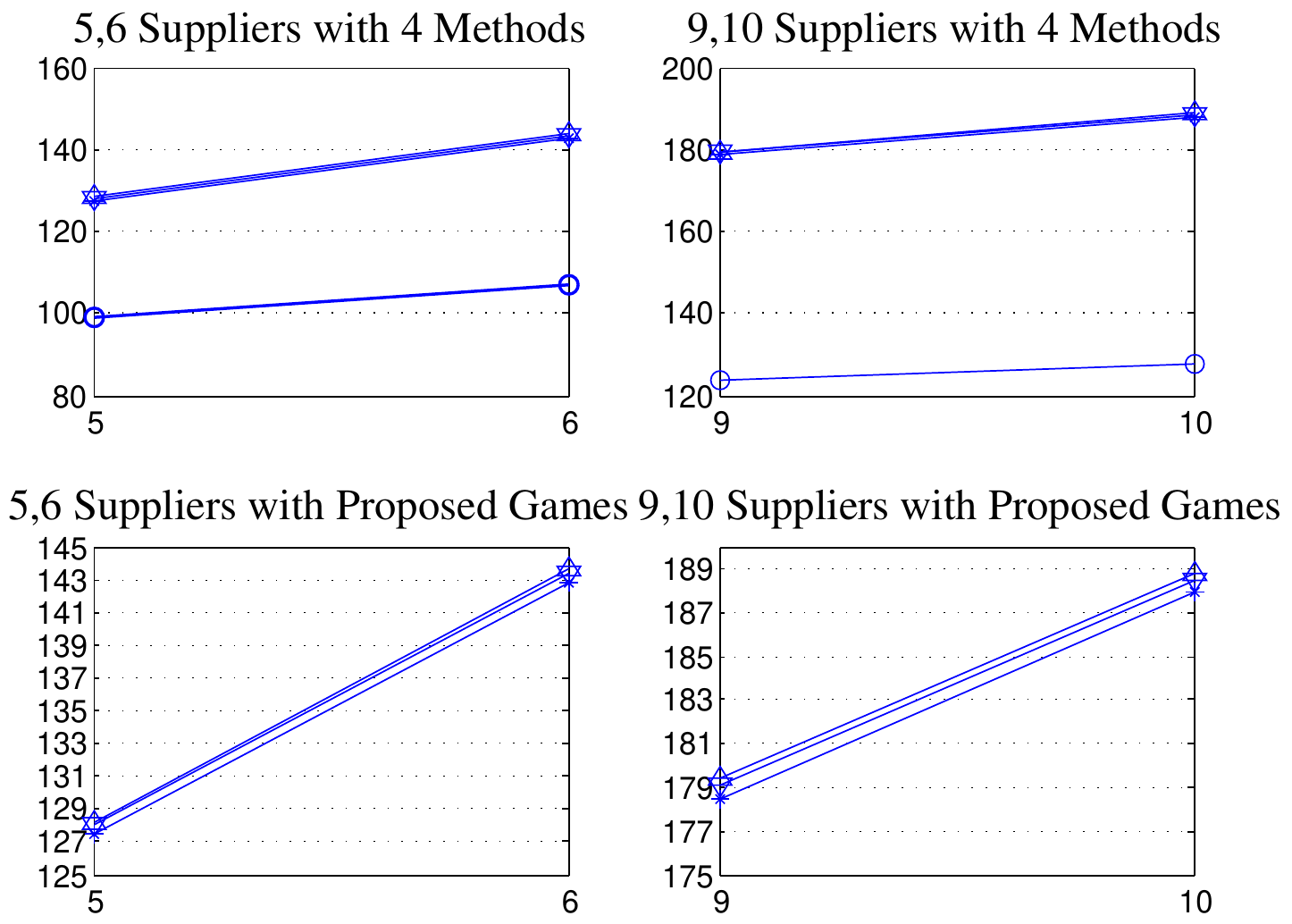}}
\end{center}
\caption{Comparison of total selling volume of energy with the two game theoretic approaches we proposed.}
\label{fig-eg7}
\end{figure}
The total selling volume through different cooperation method is shown in Figure 6. The Stackelberg Model based Game 1 is the case: $m=1,n=\{1, 2, \dots, 19\}$, i.e., there are one supplier first moving to choose action and the number of the second moving supplier increases from $1$ to $19$. The Stackelberg Model based Game 2 is the case: $m=\{1, 2, \dots, 19\}, n=1$, i.e., the number of the first moving supplier increases from $1$ to $19$, and there is one supplier second moving to choose action. As we see from Figure 6, with the same number of energy suppliers to sell their excess energy, the total selling volume of energy through the Static Energy Cooperation Method is far less than that though the game theoretic approach. It illustrates that, energy cooperation through the game theoretic approach can highly improve the utilization ratio of the harvested energy distributed in the energy harvesting WSN. In addition, the more sensor nodes in the game to supply energy the more surplus harvested energy will be sold to the energy insufficient node. The total selling volume of the Stackelberg Model based Game and Cournot Model based Game is similar, however the performance is still different. Zooming up Figure 6 and freely choosing a part, take $5$, $6$ suppliers and $9$, $10$ suppliers as examples shown in Figure 7. The Stackelberg Model based Game sells more energy than the Cournot Model based Game. It means that in the aspect of increasing the utilization ratio of the harvested energy, the dynamic game we propose is better than the static game. More specialized, the Stackelberg Model based Game 2 sells more energy than the Stackelberg Model based Game 1. This reflects the first moving advantage of the Stackelberg Model based Game.
\begin{figure}
\begin{center}
\scalebox{0.5}{\includegraphics[width=16cm,height=12cm]{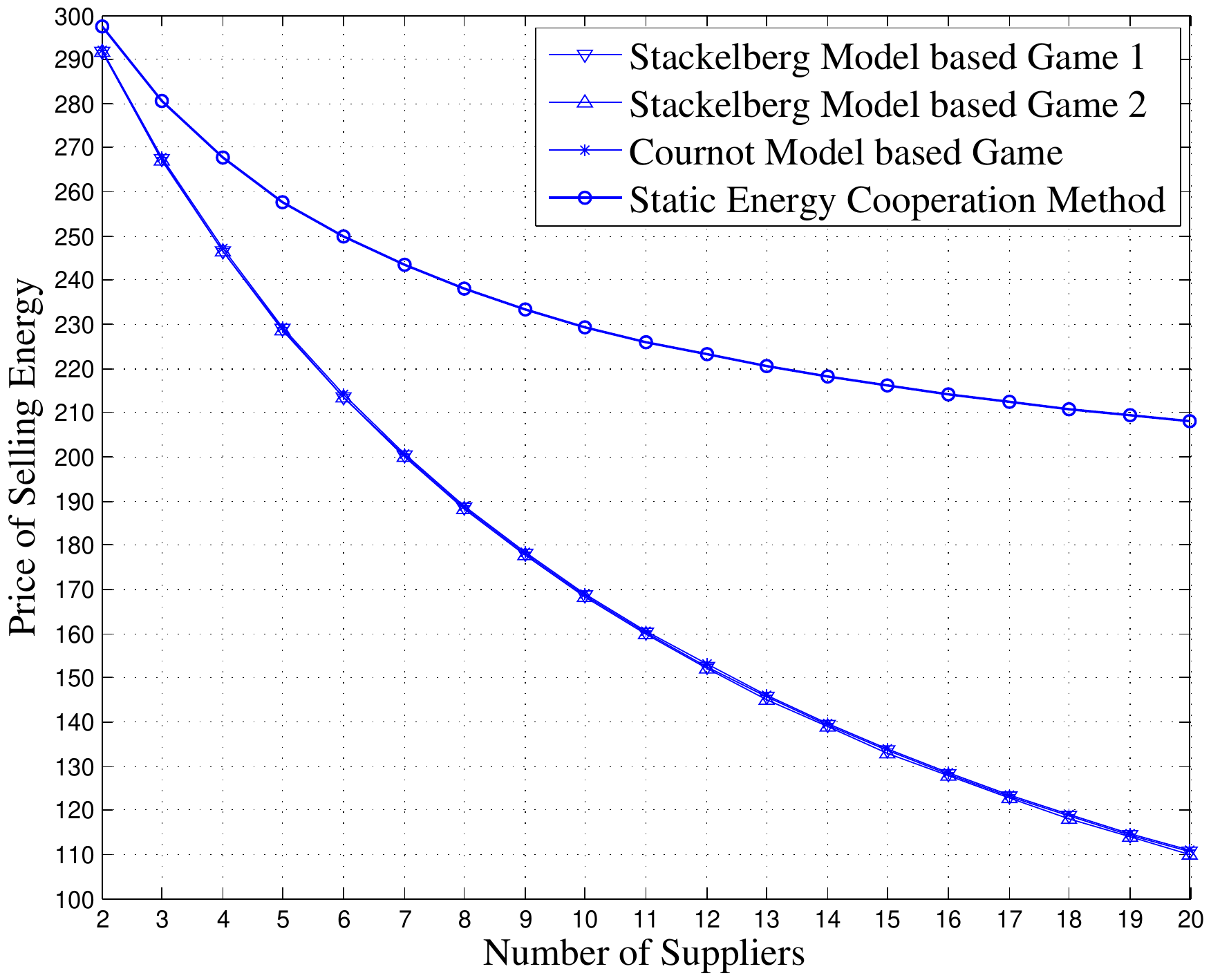}}
\end{center}
\caption{Energy price with different energy trading method.}
\label{fig-eg8}
\end{figure}

The price of the selling energy is shown in Figure 8. With the increase of the number of energy suppliers, the price of selling energy through the game theoretic approach is much lower than that through Static Energy Cooperation Method. It is expected, since the more energy is in the market to be sold the price of the energy will be lower. We have verify in Figure 6 that energy supplier node will provide more energy to the demander node through the game theoretic approach. The price of energy sold through the Stackelberg Model based Game is lower than that through the Cournot Model based Game. The first moving advantage of the Stackelberg Model based Game is also reflected in the price. The lowest price of energy appears in the Stackelberg Model based Game 2.

\section{Conclusions}
In this paper, we proposed a game theoretic framework for energy cooperation in wireless sensor networks with energy harvesting and wireless power transfer. Based on the optimal inventory amount of energy at each sensor node, sensor nodes with excess energy sold part of their energy to nodes with energy shortage through the Stackelberg Model based Game and Cournot Model based Game we designed to balance the energy at each sensor node and increase the total energy utilization ratio. The numerical results showed that compared with the static energy cooperation method, energy cooperation through the game theoretic approach can highly improve the utilization ratio of the harvested energy distributed in the energy harvesting WSNs by a higher selling volume of energy with a lower price. The Stackelberg Model based Game sold more energy than the Cournot Model based Game, i.e., the dynamic game was better than the static game. More specialized, the Stackelberg Model based Game 2 sold more energy than the Stackelberg Model based Game 1. This reflected the first moving advantage of the Stackelberg Model based Game.

\section{Acknowledgments}

The author would like to thank the editor and reviewers for their detailed reviews and constructive comments, which have helped to improve the quality of this paper. This work has been supported by the National Natural Science Foundation of China (61571059).


\appendix

\end{document}